\documentclass[12pt]{iopart}
\usepackage{graphicx} 
\usepackage{xcolor}
\usepackage{amssymb}
\makeatletter
\@namedef{ver@amsmath.sty}{}
\makeatother
\usepackage{amsmath}
\newcommand\diff{\mathrm{d}}

\usepackage[normalem]{ulem}

\usepackage{epstopdf}
\begin{document}

\title{Random motion of a circle microswimmer in  a random environment}

\author{Oleksandr Chepizhko \& Thomas Franosch}

\address{Institut f\"ur Theoretische Physik, Universit\"at Innsbruck, Technikerstra{\ss}e 21A, A-6020 Innsbruck, Austria.}
\ead{oleksandr.chepizhko@uibk.ac.at}
\vspace{10pt}
\begin{indented}
\item[]May 2020
\end{indented}

\begin{abstract}
We simulate the dynamics of a single circle microswimmer  exploring a  disordered array of fixed obstacles.
 The interplay of  two different types of randomness, quenched disorder and stochastic noise, is investigated to unravel their impact on the transport properties. 
We compute lines of  isodiffusivity as a function  of the rotational diffusion coefficient and the obstacle density.  
 We find that  increasing noise or disorder tends to amplify  diffusion, yet for large randomness the competition leads to a strong suppression of transport.
 We rationalize both the suppression and amplification of transport by comparing the relevant time scales of the free motion to the mean period between collisions with obstacles.
\end{abstract}

\section{Introduction}

Transport properties of active particles~\cite{Romanczuk2012,Elgeti2015} change significantly when they are exposed to a strongly heterogeneous medium~\cite{BechingerReview2016,Reichhardt2017review}. 
Both amplification~\cite{Makarchuk2019,Chepizhko2019,Jakuszeit2019} and suppression of diffusion~\cite{Chepizhko2013,Zeitz2017,Morin2017pre,SosaHernandez2017,Frangipane2019} with an increase of introduced obstacle density has been found in various scenarios, in experiments and in computer simulations. 
For active particles, not only is diffusion affected, but  ratchet effects~\cite{Reichhardt2017review}, negative differential mobility~\cite{Benichou2014,Reichhardt2018JPCM,Benichou2018}, and clogging~\cite{Peter2018} emerge.
Similar peculiar behavior has been seen for active Janus particles 
in visco-elastic media, which has been rationalized in terms of retarded torques coupling back to the propulsion force~\cite{Lozano2018,Narinder2018}, or for active   particles exposed to external fields as in gravitaxis~\cite{tenHagen2014} and chemotaxis~\cite{Friedrich2008,Kaupp2016}.

When these active particles undergo scattering from the inhomogeneities in the environment, diffusion  is usually suppressed~\cite{Zeitz2017,Jakuszeit2019,BrunCosmeBruny2019}.
Yet,  transport may be also enhanced, particularly, for circle microswimmers~\cite{Reichhardt2013}. 
In general, active particles can interact with their surroundings in complex ways, for example, the microswimmers can follow the boundaries of obstacles, sometimes for particularly long times~\cite{Denissenko2012,Takagi2014,Nosrati2015,Brown2016,Wykes2017},
which can be rationalized using hydrodynamic theories~\cite{Lauga2006,Berke2008,Spagnolie2015,Kuron2019}. 
On one hand, the boundary-following mechanism can slow diffusion~\cite{Brown2016,SosaHernandez2017}, if the particles are trapped for long times around heterogeneities.
On the other hand, there is theoretical~\cite{Bertrand2018}, numerical~\cite{Kamal2018,Chepizhko2019,Jakuszeit2019}, and experimental~\cite{Makarchuk2019,Creppy2019} evidence that adding  obstacles can, under certain conditions,  speed up transport. For example,  diffusion is amplified if the microswimmers are scattered forward~\cite{Jakuszeit2019}, or simply allowed to propagate along connected obstacles~\cite{Chepizhko2019}. Yet, in more general cases the properties that define the distinction between  enhancement and  hindrance of transport still need to be investigated.

Up to now a significant number of theoretical and experimental studies have used non-overlapping obstacles or pillars placed on a regular lattice~\cite{Brown2016,Wykes2017,Chamolly2017,Jakuszeit2019,BrunCosmeBruny2019}, while only few studies~\cite{Zeitz2017,Morin2017pre,SosaHernandez2017} focused on random environments. Typically, a probe particle has been chosen to be a straight-swimming active Brownian particle~\cite{Chepizhko2013,Zeitz2017}, or a particle undergoing run-and-tumble dynamics~\cite{Sandor2017,Bertrand2018}. 
To extend our understanding of transport properties of  real systems,  both these constraints should be relaxed. The paradigmatic 
Lorentz model~\cite{Lorentz1905,Bauer2010,Mandal2017} constitutes a reference system to study how  transport properties of probe particles in heterogeneous media depend on the microscopic motion  of the particles and their environment~\cite{Schnyder2015,Spanner2013,Hofling2013,Petersen2019}. 
The main feature of the Lorentz model is that the obstacles are placed randomly and can overlap. 
Regarding the model of an active particle, a Brownian \emph{circle} swimmer can be considered as a more realistic approach than those used previously, as, in general, microswimmers will not move in straight lines even for short times. 
Rather, the trajectories will be intrinsically curved, due to asymmetries in shape or the propulsion mechanism~\cite{Kummel2013,Utada2014}, or hydrodynamic coupling to walls~\cite{Lauga2006,PerezIpia2019}. 
If the angular drift is large compared to the rotational diffusion, many circles are completed before the orientation is randomized~\cite{vanTeeffelen2008, Kurzthaler2017}. 
Currently, a complete study of transport amplification and suppression  which includes both circular motion subject to noise and the wall-following mechanism, in a randomly distributed array of obstacles, is lacking. 

Here, we investigate the dynamics of a realistic model for a circle microswimmer in a disordered environment. 
 We start by adding rotational diffusion  to the motion of an ideal active circle swimmer~\cite{Chepizhko2019}. 
We show that  a small  angular noise slightly amplifies diffusion relative to an ideal  microswimmer and leaves the dependence on its orbit radius almost untouched. 
For  high values of noise the diffusivity becomes independent of the orbit radius and is determined solely by the obstacle density.
Then, we construct isodiffusivity contours in the non-equilibrium state diagram spanned  by the density of obstacles and the rotational diffusion coefficient.
We show that small amounts of both kinds of randomness amplify  diffusion, while their interplay at large values leads to a suppression of transport.
The position of the boundary between  regions of enhanced and hindered diffusion strongly depends on the orbit radius.
We explain the differences in the transport by exploring  the short-time behavior of the mean-squared displacement of a free  noisy microswimmer.

\section{Model and methods}

We consider a circle microswimmer meandering in a disordered array of disk-like obstacles in a plane. 
In free space the  microswimmer moves with fixed propulsion speed $v_0$ along its instantaneous orientation, parameterized in terms of a time-dependent angle $\theta(t)$. Then, the particle moves with velocity 
\begin{equation}\label{eq:rdot}
    \frac{\diff }{\diff t} \vec{r}(t)= v_0 (\cos\theta(t),
    \sin \theta(t) )^T  \,.
\end{equation}
The orientation itself changes in time by an angular drift $\Omega$ as well as by rotational diffusion~\cite{vanTeeffelen2008,Kummel2013,Kurzthaler2017}
\begin{equation}
    \frac{\diff}{\diff t} \theta(t) = \Omega + \sqrt{2 D_\theta}\, \xi(t) \,.
\end{equation}
The direction of motion experiences a constant  drift $\Omega=v_0/R >0$ (particle moves counterclockwise), where $R$ is referred to as the orbit radius, and the orientational dynamics is subject to Gaussian white noise  $\xi (t)$, characterized by $\langle \xi (t) \rangle = 0$, $\langle \xi (t) \xi(t') \rangle = \delta(t-t')$  with rotational diffusion coefficient $D_\theta> 0$.  In principle, one could add to Eq.~(\ref{eq:rdot}) also translational noise with short-time diffusion coefficient $D_{\textrm{\scriptsize  trans}}$. 
Yet, for microswimmers active propulsion typically dominates translational diffusion (except at short times) therefore we set $D_{\textrm{\scriptsize trans}}=0$ in the following, in line with other works on active Brownian particles~\cite{BechingerReview2016,Basu2019}, to focus on the transport properties purely caused by an interplay of rotational diffusion with quenched disorder.  
We integrate the equations of motion by means of event-driven (pseudo-) Brownian dynamics simulations~\cite{Scala2007}.

\begin{figure}
  \centering
  \includegraphics[width=0.2\textwidth]{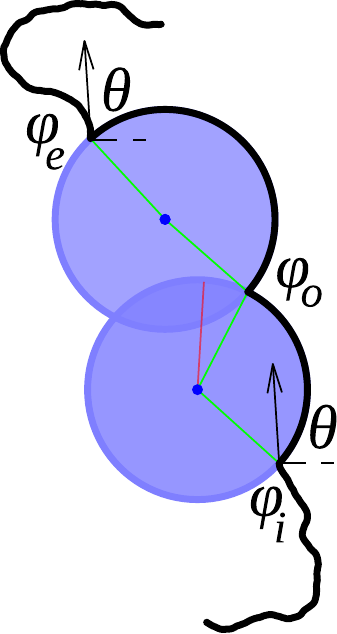}
  \caption{Illustration of the particle interaction with obstacles. 
  The particle moves from bottom to top. The trajectory of the particle is shown as solid black line. The obstacles are shown as blue circles. The particle's orientation $\theta$ remains unchanged throughout its motion along the obstacle boundaries.  The position of the collision on the first obstacle is denoted as $\varphi_i$, $\varphi_o$ is the position of the intersection of obstacles,  $\varphi_e$ refers to  the escape position on the second obstacle. The angles $\varphi$ are measured with respect to the corresponding obstacle. Green lines are radii connecting centers of obstacles with corresponding particle positions. The red line denotes the escape position that was not possible to reach because of obstacle overlap.
  In the given example a particle collides with an obstacle at the point characterized by the angle $\varphi_i$.  Then a random number is drawn and a new escape position is calculated as  shown by the red line.  Yet, the position is not reachable as the obstacles overlap.  Thus, the particle moves to the position of obstacle intersection, $\varphi_o$.  
  Next, a new random number is drawn and a new escape position $\varphi_e$  is computed.  As  the  position  is  not  inside  another  obstacle  the  particle  leaves  the  surface  into  accessible space.
  }
  \label{fig:illustration_of_interaction}
\end{figure}

We refer to  our particle as a ``microswimmer'' since the interaction with the obstacles is hydrodynamic in nature, not steric, as it usually is for active Brownian particles. 
However, we will model this interaction with a simplified rule.
The microswimmer interacts with obstacles \emph{via} a boundary-following mechanism recently introduced~\cite{Chepizhko2019}. 
Upon hitting an obstacle at a polar angle $\varphi_i$ (relative to the center of the obstacle) the microswimmer starts to follow its boundary~\cite{Chepizhko2019}, see Fig.~\ref{fig:illustration_of_interaction}. 
The orientation of the swimmer $\theta$ remains fixed during this interaction process. 
A random number $\Delta \in [ -0.9\pi/2, 0.9\pi/2 ]$ is drawn and the escape position on the boundary $\varphi_e$ is computed as follows,
\begin{equation}
    \varphi_e = (\theta - \Delta) \, \mathrm{mod} \,2\pi\,.
 \end{equation}
%
%
The escape position $\varphi_e$ does not  depend explicitly on the collision position $\varphi_i$. The choice of the interval for $\Delta$ assures that the direction of motion $\theta$ will point outside of the obstacle when the microswimmer reaches the position $\varphi_e$.
If the microswimmer cannot reach this position because it hits another obstacle first (i.e. when an intersection with the next obstacle is at $\varphi_o$, $\varphi_i<\varphi_o<\varphi_e$), a new random number is drawn and the process continues until the microswimmer escapes from the surface of the connected cluster of obstacles. 
We illustrate the interaction in Fig.~\ref{fig:illustration_of_interaction}.

The rules have been chosen to fulfill the following constraints.
The random angle $\Delta$ should not come too close to $\pm \pi/2$ since this leads to numerical artifacts, the particles get trapped and many recollisions may occur. 
 We choose the particle's orientation $\theta$ to be constant while sliding along the obstacle, rather than evolving according to some noisy dynamics, in order  to have a closer connection to the previous idealized model~\cite{Chepizhko2019}. Also, this rule simplifies the numerical simulations while being sufficiently close to  real systems.

The environment consists of randomly and independently  placed obstacles of size $\sigma>0$, which serves as the unit of length. Similarly  $\sigma v_0^{-1}$ sets the unit of time.
Then, the structural properties of the obstacle configuration are characterized solely by the reduced density 
\begin{equation}
    n^* := \frac{N \sigma^2}{L^2}\,,
\end{equation}
 a dimensionless parameter, with $N$ being the number of obstacles and $L$ the linear system size.
 The free motion can be characterized by another dimensionless parameter, referred to as the quality factor 
 \begin{equation}
     M := \frac{\Omega}{2\pi D_\theta} = \frac{v_0}{2 \pi R D_\theta}\,.
     \label{eq:M_defenition}
 \end{equation}
Then $M$ is a measure of how many circles a microswimmer can complete before the orientation becomes randomized 
in  diffusion time $1/D_\theta$~\cite{vanTeeffelen2008,Kurzthaler2017}.

In our system, long-range transport prevails  at any obstacle density smaller than the critical one $n^*_c=0.359081...$~\cite{Hoefling2006} where  the localization  transition occurs. To characterize the transport properties of the system at  densities $n^*<n^*_c$ we have measured the mean-squared displacement \begin{equation}
\delta r^2(t) := \left\langle [ \vec{R}(t) - \vec{R}(0) ]^2 \right\rangle,   
\end{equation}
where $\vec{R}(t)$ denotes the position of the swimmer at time $t$. 
At long times the mean-squared displacement is expected to be proportional to time
\begin{equation}
    \delta r^2(t) \simeq 4 D t\,, \quad t\to \infty,  
\end{equation}
with the diffusion coefficient (diffusivity) $D$.~In our data analysis we extract the diffusivity by the limit 
\begin{equation}
    D=  \lim_{t \rightarrow \infty} \frac{1}{4} \frac{\diff}{\diff t} \delta r^2(t).
\end{equation}

In the absence of obstacles the long-time diffusion coefficient   can be calculated analytically~\cite{vanTeeffelen2008,Kurzthaler2017,Ebbens2010} 
\begin{equation}
D=    D (R, D_\theta,n^* = 0) = \frac{v_0^2 D_\theta}{2 (D_\theta^2+v_0^2/R^2)}.
    \label{eq:D_vs_params_nstar0}
\end{equation}
One infers that for each value of $R$ there is an optimal rotational diffusion coefficient $D_\theta^\textrm{\scriptsize opt}$  maximizing transport in the obstacle-free system:
\begin{equation}
    D_\theta^\textrm{\scriptsize opt} := \frac{v_0}{R}\,=\Omega.
    \label{eq:Dtheta_opt}
\end{equation}
Together with the definition from Eq.~(\ref{eq:M_defenition}) we obtain the universal value of the quality factor that maximizes diffusion,
    $M^\textrm{\scriptsize opt} := 1/2\pi$.

Our reference system will be the ideal circle microswimmer ($D_\theta = 0$) in a crowded environment. There the state diagram   consists of three regions~\cite{Chepizhko2019} separated by sharp boundaries. Essentially, it is the same as the corresponding state diagram of the magneto-transport problem~\cite{Kuzmany1998,Schirmacher2015} for electrons moving in a constant perpendicular magnetic field and interacting with obstacles via \emph{specular scattering}. At low obstacle 
 densities an   \emph{orbiting} state emerges. Here  the orbit radius is not large enough for a microswimmer to reach one  obstacle cluster from another, such that  the microswimmers are simply localized around a finite number of obstacles, and so there is no diffusion.  
 Upon   increasing  the density, the system undergoes a \emph{meandering} transition to the  \emph{diffusive} state, where transport of the microswimmers through the whole system occurs.
 %
The  meandering transition depends on the obstacle density and the  orbit radius  as 
\begin{equation}
    n_m^* (\sigma/R) = n_c^* \frac{(\sigma/R)^2}{(\sigma/R+1)^2}\,,~n^*_c=0.359081\dots
    \label{eq:meandering_transition}
\end{equation}
This has been rationalized~\cite{Kuzmany1998,Schirmacher2015} by an underlying percolation transition of disks made of obstacles and 'halos' thus associating an effective radius  $\sigma+R$ to each obstacle.
Last, at densities $n^*>n^*_c$ a  \emph{localized} state emerges since the void space between the obstacles ceases to percolate, such that the microswimmers are trapped in separated pockets of void space, resulting in no long-range transport.

Here we  explore the state diagram of our system and determine the diffusivity $D$ for the noisy circle microswimmer in crowded environments.  
To illustrate our results we have computed  contours of  isodiffusivity, based on the data  obtained from simulations. 
The linear system size $L=10^4\sigma$ is such that finite size effects are irrelevant and computer simulation time is reasonable.
The total simulation time is $10^5-10^7 \sigma v_0^{-1}$. 
For each data point an average is performed over $200-400$ disorder realization. 

\section{Results}
First, we consider how adding  rotational noise to the dynamics of an ideal microswimmer smears the meandering transition~\cite{Chepizhko2019} and discuss the changes of the isodiffusivity lines.
%
Next, we study the interplay of the  two kinds of randomness, the dynamic angular noise and the quenched  configurations of obstacles for different  values of the orbit radius $R$. %
We highlight regions of the state diagram where  diffusion is amplified and where it is suppressed.

\begin{figure*}[th!]
  \centering
  \includegraphics[width=0.99\textwidth]{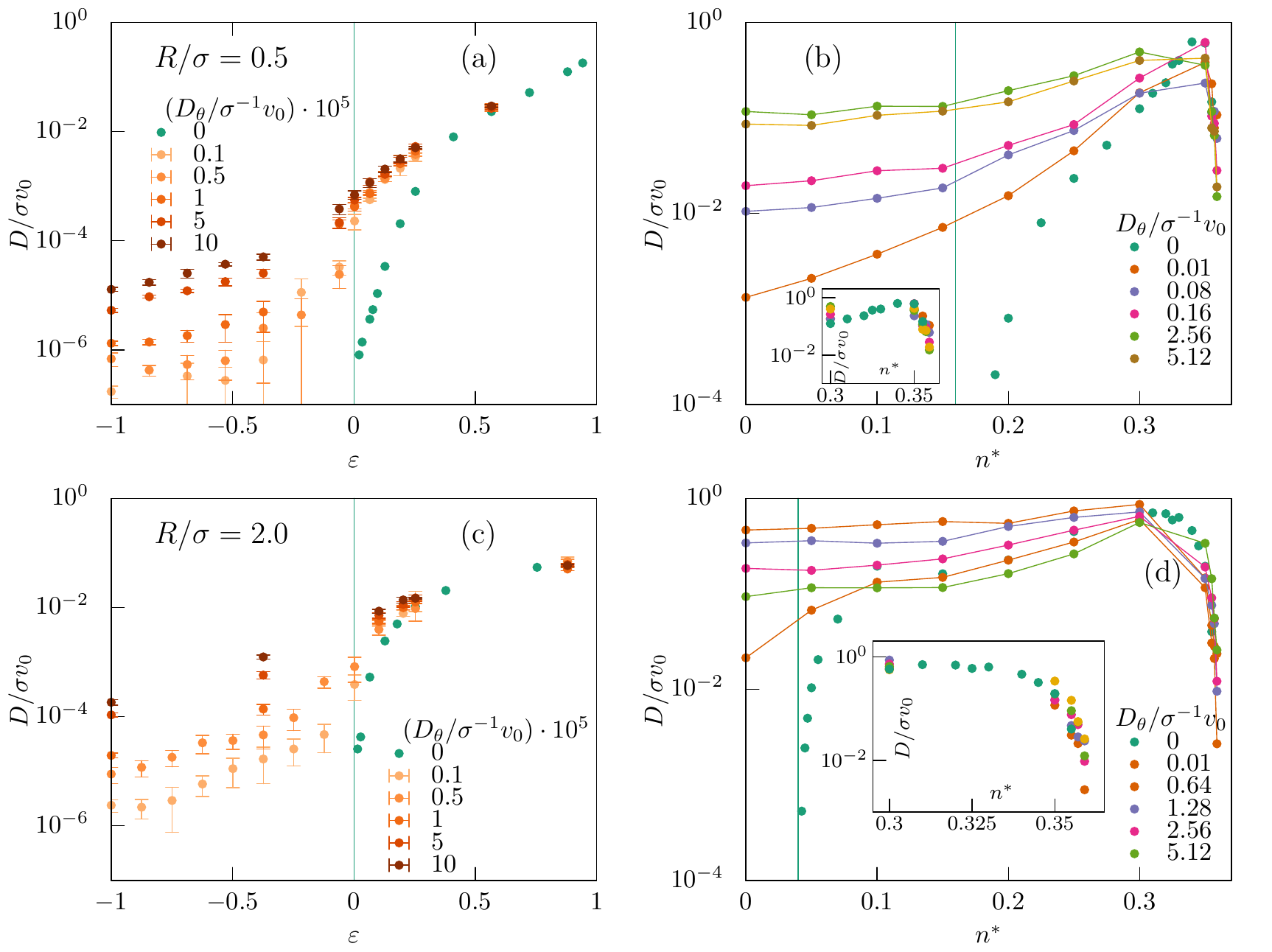}
  \caption{Diffusivity $D=D(R,D_\theta,n^*)$ vs. density of obstacles $n^*$ (or separation parameter $\varepsilon=(n^*-n^*_m(R))/n^*_m(R)$) for orbit radius $R=0.5 \sigma$ (a,b) and $R=2.0 \sigma$ (c,d)  for increasing values of the rotational diffusion $D_\theta$.~Vertical line shows the density $n^*_m(R)$ corresponding to the meandering transition, computed from Eq.~(\ref{eq:meandering_transition}). ~Panels (a) and (c) focus on the former \emph{meandering} transition showing the data for small separation parameters $\varepsilon$ and at relatively low angular noise.~Panels (b) and (d) show the diffusivities over the full density range and at higher values of $D_\theta$.~Insets in panels (b,d) show  close-ups of the diffusivity at high obstacle densities.
  Data  for ideal microswimmer ($D_\theta=0$) taken from Ref.~\cite{Chepizhko2019}.
  }
  \label{fig:smearing}
\end{figure*}

\begin{figure*}[th!]
  \centering
  \includegraphics[width=0.99\textwidth]{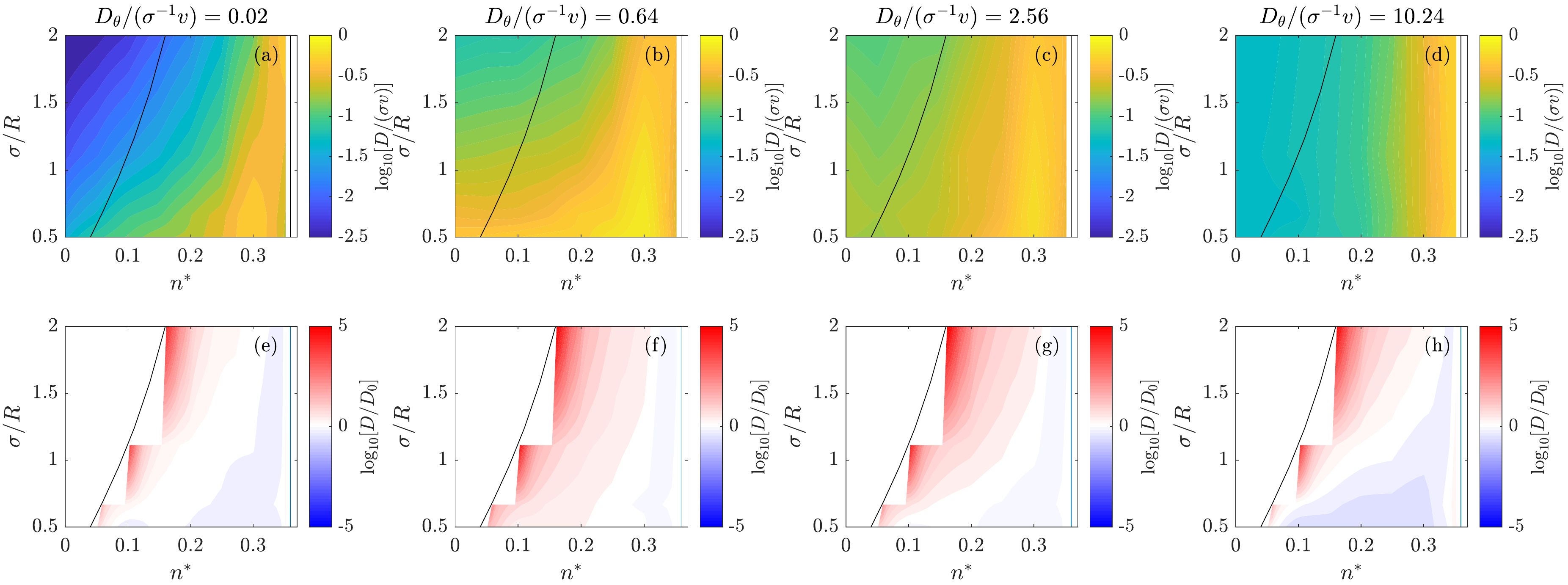}
  \caption{(a)-(d): Isodiffusivity contours $D=D(R,D_\theta, n^*)$, and (e)-(h):  ratios of diffusivity to the diffusivity of the idealized model $D(R,D_\theta, n^*))/D(R,D_\theta=0, n^*)$ in the state diagram spanned by obstacle density $n^*$ and inverse orbit radius $\sigma/R$ for increasing values of angular noise $D_\theta /\sigma^{-1} v_0$: (a),(e) $0.02$, (b),(f) $0.64$, (c),(g) $2.56$, (d),(h) $10.24$. 
  In (a)-(d) color coding corresponds to $\log_{10} (D / \sigma v_0) $. In (e)-(h) red color corresponds to amplification whereas blue indicates suppression of diffusion.  
  The  black curve indicates  the meandering transition of the ideal microswimmer model~\cite{Chepizhko2019},  Eq.~(\ref{eq:meandering_transition}).  Note that to the left of the meandering transition for the ideal circle swimmer the diffusion coefficient evaluates to  zero.
  The vertical black line shows the percolation transition at the density value $n_c^*= 0.359\ldots$.  In  the panels of this figure, $\sigma/R$ is used to facilitate the comparison with the ideal model, Ref.~\cite{Chepizhko2019} and magneto-transport model, Ref.~\cite{Schirmacher2015}.
 }
  \label{fig:isodiff_diagrams_Dthetas}
\end{figure*}

\subsection{State diagram and isodiffusivity contours}

For  any nonzero angular noise  the meandering transition is smeared and the orbiting state is no longer present. Particles can reach  different obstacle clusters by  random reorientations, and this  will always happen provided one waits long enough. 
The smearing of the phase transition is illustrated in  Fig.~\ref{fig:smearing}(a,c) in terms of the diffusivity $D$ as the density is increased towards the ideal meandering transition. 
The distance to the transition is characterized by a separation parameter $\varepsilon=(n^*-n^*_m(R))/n^*_m(R)$.
Indeed,  transport occurs now at any obstacle density due to the orientational noise.  
Yet, by lowering  the angular noise  $D_\theta$ the translational diffusion $D(R,D_\theta,n^*)$ approaches the case of the ideal microswimmer. 
Thus, the smearing of the transition is a continuous process and
the results of the ideal microswimmer~\cite{Chepizhko2019} remain valid for sufficiently small noise. The rate of convergence depends on the orbit radius. For small orbit radii [Fig.~\ref{fig:smearing}(a)] the convergence rate is slightly slower then for the larger ones [Fig.~\ref{fig:smearing}(c)]. This is due to the higher value of the obstacle density of the meandering transition for smaller orbit radii. 

It should be mentioned that if translational diffusion was included, the meandering transition would also be smeared even in the absence of rotational noise. 
However, for realistic microswimmers the rotational diffusion is anticipated to be the main effect and the additional smearing due to translational diffusion should be small.

Now we zoom out from the meandering transition and consider the whole density range~Fig.~\ref{fig:smearing}(b,d). For the intermediate obstacle densities the diffusivity $D$ increases when the angular noise is increased up to its optimal value for a small orbit radius Fig.~\ref{fig:smearing}(b) and a large one Fig.~\ref{fig:smearing}(c). At larger $n^*$, closer to the percolation transition, transport is dominated by the obstacles, in particular, by the crowding-enhancement transport mechanism, while angular noise is of minor importance. This can be seen in detail in the insets in Fig.~\ref{fig:smearing}(b,d). 
However, we note that the maximum diffusivity for each fixed curvature almost never exceeds the maximum value of the diffusivity in an ideal system, Fig.~\ref{fig:smearing}(b,d).


To get a more broad view we plot isodiffusivity contours in the state diagram spanned by the  obstacle density $n^*$ and the  orbit curvature $\sigma/R$, as in the case of the ideal circle microswimmer~\cite{Chepizhko2019} or the magneto-transport problem~\cite{Schirmacher2015}. The diagrams are presented
for increasing values of angular noise $D_\theta$, see Fig.~\ref{fig:isodiff_diagrams_Dthetas}(a-d). In the second row, Fig.~\ref{fig:isodiff_diagrams_Dthetas}(e-h), the ratio of the diffusivity to its value for the idealized system ($D_\theta=0$) is shown.
For small noise $D_\theta$, the diffusivity in the region above the meandering transition, $n^*>n^*_m(R)$ is very similar to the ideal case  [Fig.~\ref{fig:isodiff_diagrams_Dthetas}(a)]. 
Only very close to the percolation transition there is a sharp drop of the diffusivity since swimmers become trapped in isolated pockets. In most of the parameter space corresponding to the diffusive state of the ideal circle swimmer, the values of the diffusivity $D$ are again similar for both models.
To the left of the meandering transition line, i.e. in the orbiting state, this ratio remains undefined [Fig.~\ref{fig:isodiff_diagrams_Dthetas}(e)]. Away from it, in most of the diffusive state, the light shading of the color coding indicates that the diffusivity there 
is very similar to the idealized case [Fig.~\ref{fig:isodiff_diagrams_Dthetas}(e)]. Approaching the meandering transition of the idealized model the diffusivity rapidly decreases, yet remains nonzero [Fig.~\ref{fig:isodiff_diagrams_Dthetas}(a)] in  stark contrast to the idealized system where no diffusion in this region occurs.

For intermediate noise $D_\theta$ [Fig.~\ref{fig:isodiff_diagrams_Dthetas}(b,c)], the dependence of the diffusivity on the orbit radius is still strong even for the obstacle-free system $n^*=0$, as suggested by Eq.~(\ref{eq:D_vs_params_nstar0}). 
Upon increasing the density of obstacles the diffusivity again grows, however, the growth occurs mainly in the regime of the ideal diffusive state. For obstacle densities below the ideal meandering transition the dependence on the obstacle density is rather weak, such that the isodiffusivity lines are almost horizontal. An amplification of diffusion in the ideal diffusive state is revealed in Fig.~\ref{fig:isodiff_diagrams_Dthetas}(f,g).

For even higher angular noise  [Fig.~\ref{fig:isodiff_diagrams_Dthetas}(d)] the dependence of the diffusivity on the orbit radius fades out. 
This can be explained by the fact that the particle's free motion ceases to be circular, since the quality factor $M$ is very low. 
Nevertheless, the diffusion coefficient increases with the density of obstacles except in the  close vicinity of the percolation transition. 
In comparison to the ideal case, Fig.~\ref{fig:isodiff_diagrams_Dthetas}(h) shows that diffusion is suppressed in large areas of the parameter space.

\begin{figure*}[th!]
  \centering
 \includegraphics[width=0.99\textwidth]{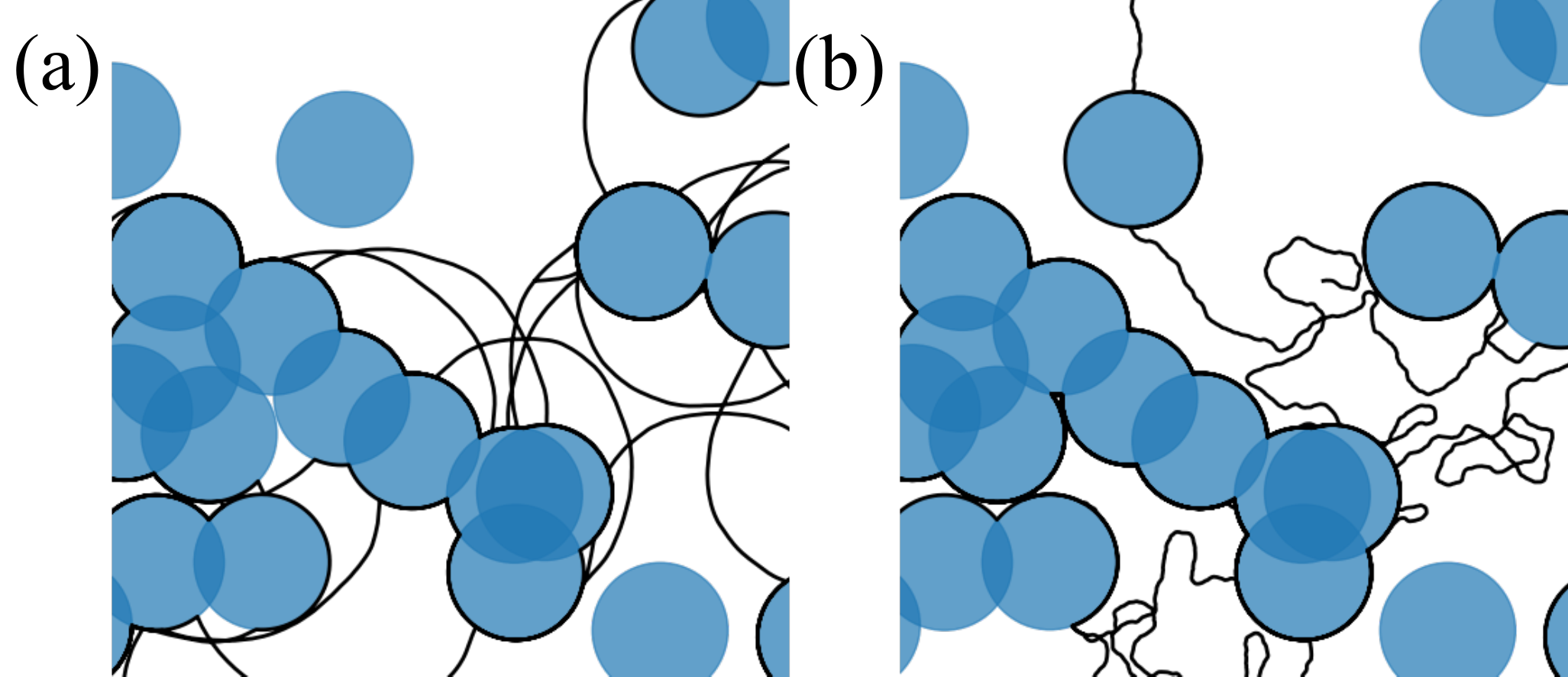}
  \caption{Trajectories of microswimmers (black) exploring the same array of obstacles (shown as blue circles) at obstacle density $n^*=0.2$ (a) low noise, $D_\theta=0.01  \sigma^{-1}v_0$ and (b) intermediate noise, $D_\theta=1.25  \sigma^{-1}v_0$.}
  \label{fig:snapshots_low_high_noise}
\end{figure*}

The qualitative difference between the low and high noise behavior becomes immediately apparent upon inspecting typical trajectories, see Fig.~\ref{fig:snapshots_low_high_noise}. At  low noise [Fig.~\ref{fig:snapshots_low_high_noise}(a)] the microswimmer trajectory in  free space is composed  of distorted circles, while for intermediate  noise [Fig.~\ref{fig:snapshots_low_high_noise}(b)],  the trajectories get significantly randomized and the noise promotes diffusion in free space. 


In summary, low values of angular noise promote transport by allowing for the possibility to reach obstacle clusters which were inaccessible in the ideal case due to a limited orbit range. At intermediate noise values, the already efficient free-space transport is enhanced by the wall-following mechanism. At very large $D_\theta$ the circular motion is so rapidly randomized that transport approaches the case of an active Brownian tracer in a disordered environment, yet the wall-following mechanism enhances transport at high obstacle densities. 

\subsection{Amplification and suppression of transport}

\begin{figure*}[th!]
  \centering
  \includegraphics[width=0.99\textwidth]{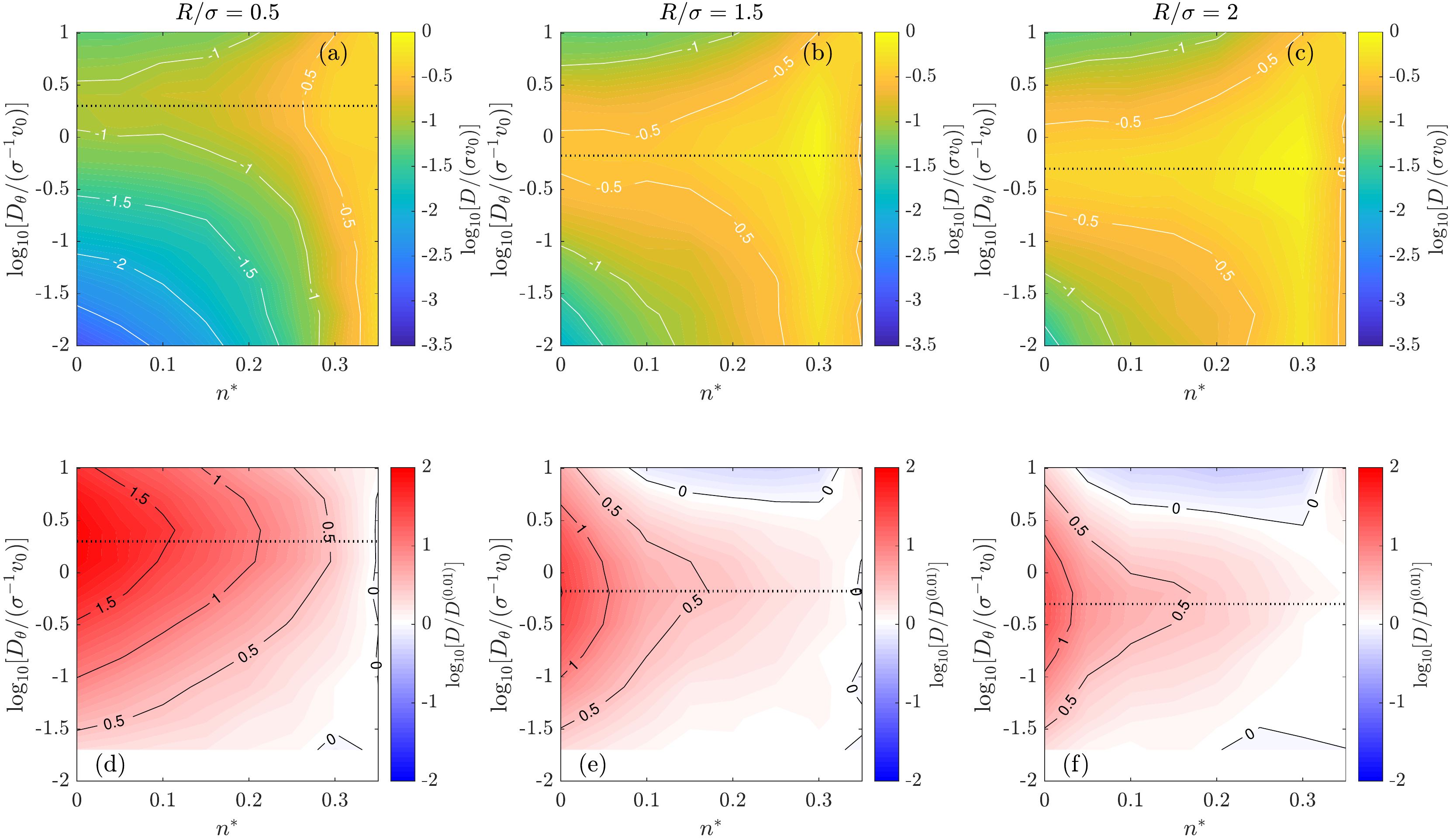}
  \caption{Panels (a,b,c) Isodiffusivity contours $D = D(R,D_\theta,n^*)$ in the plane $n^*$ -- $D_\theta$ for  three orbit radii $R/\sigma$: (a,d) $R/\sigma=0.5$, (b,e) $R/\sigma=1.5$, (c,f) $R/\sigma=2.0$.  Color coding corresponds to $\log_{10}{D/\sigma v_0}$.
Panels (d,e,f):  $D/D^{(0.01)}$ Diffusion coefficient relative to the diffusion coefficient at a low value of the rotational noise, $D_\theta = 10^{-2} \sigma^{-1}v_0$. 
    Red (blue) color corresponds to amplification (suppression) of diffusion.
  Dotted black lines correspond to the value of the rotational diffusion $D_\theta^\textrm{\scriptsize opt}$ which maximizes spatial diffusion $D$ in the zero obstacle case, given by Eq.~(\ref{eq:Dtheta_opt}), and thus correspond to a quality factor value $M=1/2\pi$.}
  \label{fig:all_R_isocontours}
\end{figure*}

It is instructive to study directly the interplay of quenched disorder and dynamic noise. Therefore we redraw the isodiffusivity lines in a new state diagram spanned by the obstacle density $n^*$ and the rotational diffusion coefficient $D_\theta$ for fixed  orbit radius  $R$, see Fig.~\ref{fig:all_R_isocontours}(a-c). 
We highlight regions of amplification and suppression of diffusion by relating the computed contours  to the diffusion coefficient at a small value of the rotational noise, Fig.~\ref{fig:all_R_isocontours}(d-f). We choose a system with some rotational noise over the ideal system to have a more complete state diagram, without void regions of undefined value.

We can see that the diffusivity maps are qualitatively similar for three representative orbit radii, Fig.~\ref{fig:all_R_isocontours}(a-c).
The diffusivity continuously increases when the obstacle density is increased  up to $n^*\simeq 0.3$. 
The line of constant angular noise $D_\theta=D_\theta^{\textrm{\scriptsize opt}}$ separates two regions.
When the angular noise strength is increased up to the optimal value $D_\theta^\textrm{\scriptsize opt}$, the diffusivity grows with it. 
However, as the angular noise is increased further the diffusivity starts to  decrease for any value of the obstacle density. 
We note that for smaller radius [Fig.~\ref{fig:all_R_isocontours}(a)] the optimal density of obstacles is very close to  $n^*_c$, while for large radius [Fig.~\ref{fig:all_R_isocontours}(b,c)] the optimal density value is closer to $n^*\approx 0.3$.

Next, we consider amplification or suppression of the diffusivity with respect to its value at  low rotational noise 
\begin{equation}
    D^{(0.01)}(R,n^*):=D(R, D_\theta=0.01\,\sigma^{-1}v, n^*)\,,
\end{equation} 
plotted in Fig.~\ref{fig:all_R_isocontours}(d,e,f). 
In all three panels at $n^*=0$ we see an enhancement of diffusion $D(D_\theta)/D^{(0.01)}>1$ (indicated by the red color), which is solely due to the increase of angular noise, $D_\theta$. 
When $n^* > 0$, an increase of rotational diffusion enhances the diffusivity less and less for increasing density of obstacles. 
For the smallest radius $R=0.5\sigma$, diffusion is almost always amplified [Fig.~\ref{fig:all_R_isocontours}(d)].
For the intermediate radius $R=1.5\sigma$, a region emerges at high obstacle densities where diffusion becomes suppressed when the angular noise is increased [Fig.~\ref{fig:all_R_isocontours}(e)]. 
For the largest orbit radius $R=2.0\sigma$, the size of this region increases [Fig.~\ref{fig:all_R_isocontours}(f)].
From this we draw the conclusion that the area in parameter space covered by the suppression region increases with orbit radius.


While in general  including angular noise and obstacles into the environment amplifies particle diffusion, for large values of both perturbations transport becomes hindered.
The exact position of the boundary depends on the orbit radius $R/\sigma$, with smaller values of noise strength or obstacle density needed to cause a  suppression of diffusion for larger radii.

\section{Discussion.}
 
\subsection{Role of the Mean-Squared Displacement}

\begin{figure*}
  \centering
  \includegraphics[width=0.99\textwidth]{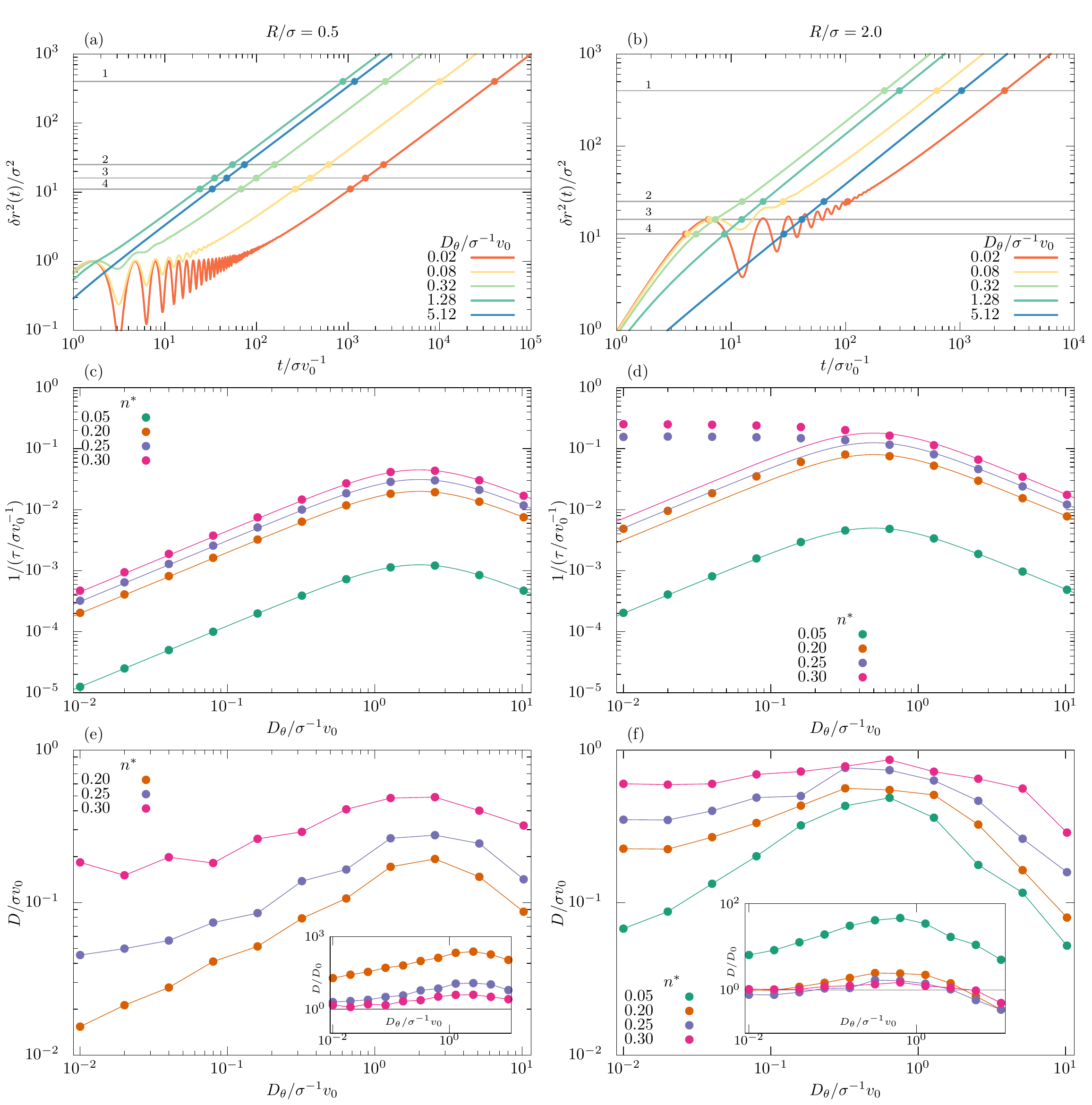}
  \caption{Mean-square displacements (curves) of a circle swimmer in an obstacle free environment for orbit radius $R=0.5\sigma$ (a) and $R=2.0\sigma$ (b) for different values of angular noise $D_\theta$, plotted from Eq~(\ref{eq:msd_for_abcs}). 
  Enumerated lines show the values of characteristic length squared $\ell^2$ for different obstacle densities, (1) $\ell^2=400\sigma^2$, $n^*=0.05$, (2) $\ell^2=25\sigma^2$, $n^*=0.2$, (3) $\ell^2=16\sigma^2$, $n^*=0.25$, (4) $\ell^2=11.1\dots\sigma^2$, $n^*=0.3$. 
  Points mark the characteristic time $\tau$, obtained by graphical (numerical) solution. 
  Inverse characteristic time $1/\tau$ as function of angular noise $D_\theta$ for $R=0.5\sigma$ (c) and $R=2.0\sigma$ (d) at several values of obstacle density $n^*$.
  Points represent graphical solutions, lines are given by Eq.~(\ref{eq:diffusive_tau}).
  Diffusivity (main panels) and ratio of diffusivity to its ideal value at given density of obstacles $D(R, D_\theta, n^*) / D(R, D_\theta=0, n^*)$ (insets) vs. angular noise $D_\theta$, for $R=0.5\sigma$ (e) and $R=2.0\sigma$ (f). Lines in panels (e,f) serve as guides to the eye. Grey line in insets marks $D/D_0=1$.}
  \label{fig:for_discussion}
\end{figure*}

In the \emph{Results} section it has been elaborated that the  amplification-suppression patterns are different for small ($R=0.5\sigma$) [Fig.~\ref{fig:all_R_isocontours}(d)] and large ($R=1.5\sigma,\,2.0\sigma$) radii [Fig.~\ref{fig:all_R_isocontours}(e,f)].
For a small radius $R/\sigma=0.5$ an amplification of transport with an increase of angular noise is observed at any obstacle density [Fig.~\ref{fig:all_R_isocontours}(d)]. It qualitatively follows the diffusivity dependence on angular noise, given by the Eq.~(\ref{eq:D_vs_params_nstar0}), where two values of angular noise, a low and high one, yield the same diffusivity value, causing similar amplification.
However, for $R/\sigma=2.0$ at intermediate-to-high densities the angular noise causes little amplification and even leads to suppression of diffusion at high noise values [Fig.~\ref{fig:all_R_isocontours}(f)]. The long-time diffusivity given by Eq.~(\ref{eq:D_vs_params_nstar0}) is not enough to explain this behavior. 

To provide an intuitive explanation for such a drastic difference we consider the mean-squared displacement of an active circle Brownian swimmer in free space~\cite{vanTeeffelen2008,Kurzthaler2017,Ebbens2010} 
\begin{eqnarray}
\delta r^2(t)  =
\frac{2 v_0^2 D_\theta^2}{(D_\theta^2+\Omega^2)^2}
\left[
 D_\theta t-1  + \frac{ \Omega^2}{D_\theta^2} (D_\theta t + 1)
  \right. \nonumber \\
+\left(1-\frac{\Omega^2}{D_\theta^2}\right) \cos(\Omega \, t) e^{-D_\theta t}  
  -\left. 
2  \frac{\Omega}{D_\theta} \sin(\Omega \, t)e^{-D_\theta t}  \right]\,.
\label{eq:msd_for_abcs}
\end{eqnarray}
The mean-squared displacement typically evolves through three different regimes: directed motion (at short times), oscillations (at intermediate times), and diffusion (at long times). 
The duration of each of the regimes depends on the time scales $\Omega^{-1}$ and $D_\theta^{-1}$.

To explain the difference in diffusivities at different $D_\theta$ for fixed $n^*$ we  consider the time needed to cover the characteristic distance between obstacles  $\ell:=\sigma/n^*$, termed the mean-free path length~\cite{Franosch2010}. 
In some cases the particle can cover a distance $\ell$ already in the regime of directed motion (large $R$, large $n^*$, low $D_\theta$). 
In other cases (high $D_\theta$, low $n^*$, low $R$) 
the diffusive regime has been entered before the length $\ell$ is reached by the microswimmer. 
The difference can be quantified by a characteristic time $\tau$, which can be  directly inferred  from the plot [Fig.~\ref{fig:for_discussion}(a,b)], as the first intersection of the mean-squared displacement curve with the line of constant $\ell^2$. 
For small radius ($R=0.5\sigma$) the mean-squared displacements  given by Eq.~(\ref{eq:msd_for_abcs}) display oscillations for low angular noise [Fig.~\ref{fig:for_discussion}(a)]. Their amplitude is small in comparison to the mean-free path $\ell$ for all  obstacle densities considered, and the intersection always occurs in the diffusive regime. 
Thus, the time $\tau$ can be computed using the long-time asymptote for the mean-squared displacement, $\delta r^2(t) = 4 D t$, as $t\rightarrow \infty$, together with Eq.~(\ref{eq:D_vs_params_nstar0}),
\begin{equation}
    \tau = \frac{\ell^2}{2 v_0^2 D_\theta}\left(D_\theta^2 +\frac{v_0^2}{R^2}\right)\,.
    \label{eq:diffusive_tau}
\end{equation}
In this case, Eq.~(\ref{eq:D_vs_params_nstar0}) can be used to describe the variation in the diffusivity.
In stark contrast, for large orbit radius ($R=2.0$)  [Fig.~\ref{fig:for_discussion}(b)] the microswimmer can access regions of distance $\ell$ while displaying circular motion. Thus, Eq.~(\ref{eq:D_vs_params_nstar0}) alone is not enough to explain the results.

The  inverse of the characteristic time $\tau$ is displayed in Fig.~\ref{fig:for_discussion}(c,d) as a function of $D_\theta$.  
For $R=0.5\sigma$, a non-monotonic behavior with a pronounced maximum for 
$1/\tau$  emerges at all densities  [Fig.~\ref{fig:for_discussion}(c)], according to Eq.~(\ref{eq:diffusive_tau}).
Yet, for $R=2.0\sigma$ beyond a  density $n^*\gtrsim 0.25$ the dependence becomes monotonically  decreasing [Fig.~\ref{fig:for_discussion}(d)] and solutions of the full equation (graphical solutions) deviate from the approximation, Eq.~(\ref{eq:diffusive_tau}), at low values of $D_\theta$.
In both cases the dependence of the diffusivity $D$ on the angular noise $D_\theta$ correlates with $1/\tau$.
For $R=0.5\sigma$ the result is not surprising, as $D$ vs. $D_\theta$ is also a non-monotonic function with the maximum at the corresponding position [Fig.~\ref{fig:for_discussion}(e)]. 
From the idealized model we know that a higher obstacle density provides a larger diffusion~\cite{Chepizhko2019}, thus the amplitude of the amplification decreases with $n^*$. 
Most importantly, the diffusivity never becomes suppressed [Fig.~\ref{fig:for_discussion}(e) inset].

For large radius $R=2.0\sigma$ the argument is more subtle [Fig.~\ref{fig:for_discussion}(f)]. 
At low densities, $n^*\approx 0.05$, the behavior is the same as for small radius.
For high densities $n^* \gtrsim 0.2$ the diffusivity remains almost constant  (similar to its value of the idealized model) as the noise is increased. It becomes slightly amplified around the optimal noise value $D_\theta^{\textrm{\scriptsize opt}}$, and
 for $D_\theta$ larger than $D^\textrm{\scriptsize opt}_\theta$ the diffusion is suppressed. 
Considering the diffusivity ratios  [Fig.~\ref{fig:for_discussion}(f) inset] only the low-density  $n^*=0.05$ is similar to  the $R=0.5$ case [Fig.~\ref{fig:for_discussion}(e) inset]. 
All other curves do not quite follow the $1/\tau$ curves at low noises. 
While the characteristic time to cover $\ell^2$ only increases with $D_\theta$ [Fig.~\ref{fig:for_discussion}(d)], the diffusivity $D$ nevertheless is amplified around the optimal angular noise $D_\theta^{\textrm{\scriptsize opt}}$ [Fig.~\ref{fig:for_discussion}(f) inset]. For larger values of angular noise the $1/\tau$ and $D$ curves decrease in a correlated way [Fig.~\ref{fig:for_discussion}(d) and (f) inset].
So, the characteristic time $\tau$ is no longer sufficient, rather we anticipate that the entire distribution of  times needed to cover an entire  range of path lengths  slightly larger than $\ell$ determines the transport properties. 
%
%
\begin{figure*}[h!]
  \centering
  \includegraphics[width=0.99\textwidth]{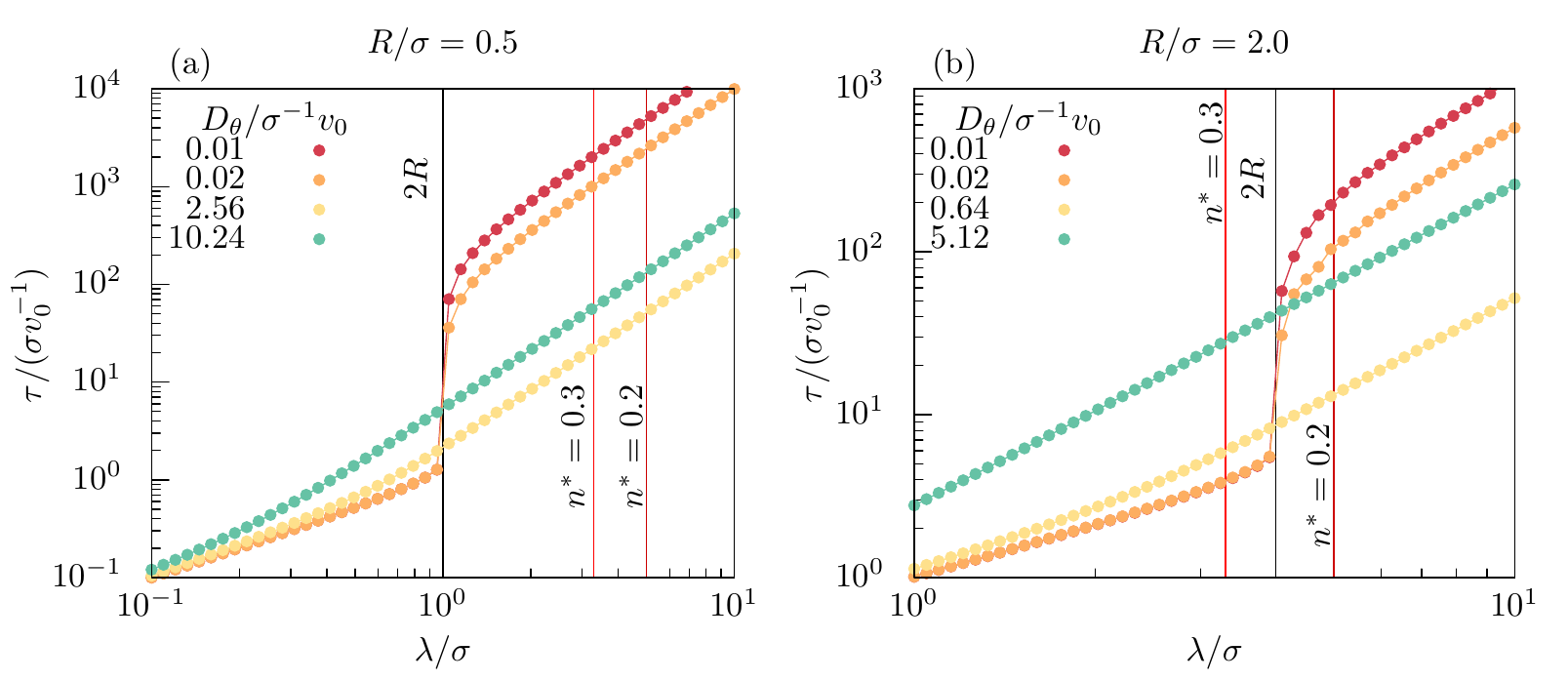}
  \caption{Characteristic time $\tau$ vs. distance $\lambda$ for various values of the angular noise $D_\theta$ for (a) $R/\sigma=0.5$ ($D_\theta^\textrm{\scriptsize opt} = 2.0\sigma^{-1}v_0$) and (b) $R/\sigma=2.0$ ($D_\theta^\textrm{\scriptsize opt}=0.5 \sigma^{-1}v_0$). Black lines indicate the diameter $2R$ and the two red lines show the characteristic lengths corresponding to obstacle density $n^*=0.20$ and $n^*=0.30$.}
  \label{fig:tau_vs_ell}
\end{figure*}

\begin{figure*}[h!]
  \centering
  \includegraphics[width=0.99\textwidth]{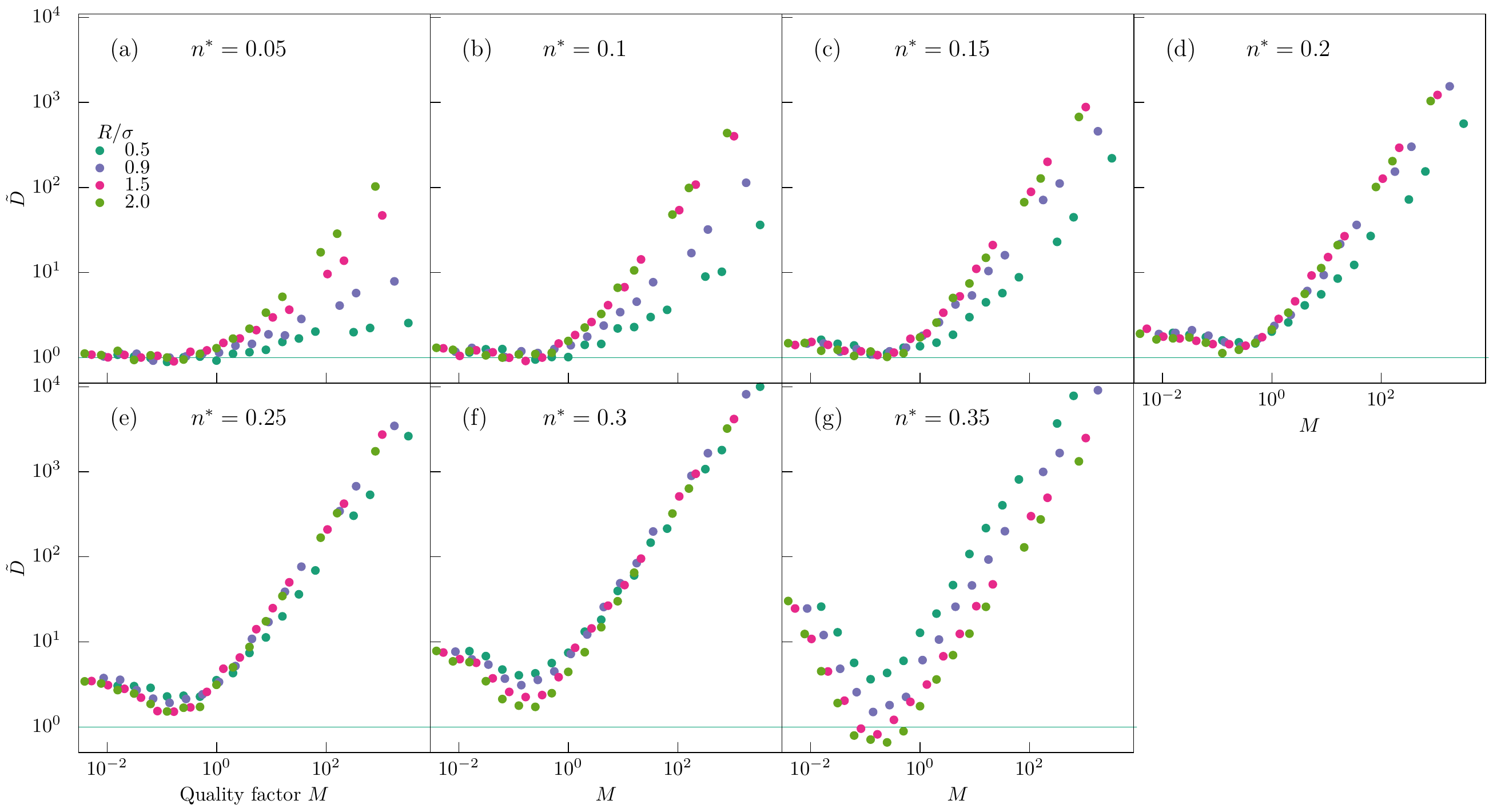}
  \caption{Rescaled diffusivity $\tilde{D} = \tilde{D}(R,M,n^*)$ [Eq.~(\ref{eq:an_equality_with_M})] as a function of  quality factor $M$. 
  Panels (a)-(g) show the dependence for increasing obstacle densities  $n^*$. Different colors represent  different values of the orbit radius.  In each panel the green line indicates $1$, which is the rescaled diffusivity in the absence of obstacles.
  }
  \label{fig:dependence_on_M}
\end{figure*}

For an ideal microswimmer the time $\tau$ to cover an arbitrary distance $\lambda$ increases until the length scale becomes equal to the orbit diameter, $\lambda = 2 R$. 
For larger values of $\lambda>2R$ the characteristic time $\tau$ does not exist, as the ideal microswimmer cannot travel farther than $2R$.
If a small amount of angular noise is introduced, length scales $\lambda>2R$ become accessible, but the time to cover them is significantly larger compared to the time to cover $\lambda=2R$.
The dependence of $\tau$ on an arbitrary length scale $\lambda$ is rather steep at low noise values in the vicinity of $\lambda \approx 2 R$, see Fig.~\ref{fig:tau_vs_ell}. 
For $R=0.5\sigma$ this steep increase occurs far away from the characteristic lengths defined by the largest considered densities [Fig.~\ref{fig:tau_vs_ell} (a)], and thus has no impact on the amplification-suppression pattern. 
For $R=2.0\sigma$ the values of $\ell(n^*)$ at $n^* \gtrsim  0.2$ are comparable with the diameter $2R$, for example, $\ell(n^*=0.25)=2R=4\sigma$, [Fig.~\ref{fig:tau_vs_ell}(b)].
The amplification of diffusion at low noises ($D_\theta \ll D_\theta^\textrm{\scriptsize opt}$) is absent because the characteristic times to cover small distances are comparable with the ones for the ideal system $\tau(\lambda<2R,D_\theta \ll 1) \approx \tau(\lambda<2R,D_\theta=0)$. Yet, times to cover slightly farther distances are quite large $\tau(\lambda>2R,D_\theta \ll D_\theta^\textrm{\scriptsize opt}) \gg \tau(\lambda<2R,D_\theta \ll D_\theta^{\textrm{\scriptsize opt}})$ and do not affect the transport properties [Fig.~\ref{fig:tau_vs_ell}(b), red and orange curves].
If the angular noise is increased up to its optimal value, $D_\theta \approx D_\theta^{\textrm{\scriptsize opt}}$, the time to cover small distances also increases, $\tau(\lambda<2R, D_\theta \approx D_\theta^{\textrm{\scriptsize opt}}) > \tau(\lambda<2R,D_\theta=0)$ [Fig.~\ref{fig:tau_vs_ell}(b), yellow curve].
However, the time to cover larger distances decreases drastically and the steep increase becomes smooth. 
%
The latter effect is more important and an amplification of transport is observed.
With a further increase of angular noise beyond its optimal value all characteristic times also increase, both for small $\lambda < 2 R$ and large $\lambda > 2R$ length scales  [Fig.~\ref{fig:tau_vs_ell}(b), aquamarine curve]. Hence, transport becomes suppressed.
This picture holds for obstacle densities below $n^*=0.3$ while the average length scale $\ell$ remains comparable to $2R$. As soon as the density becomes too low, for example $n^*=0.05$, the simple mechanism described for $R=0.5$ is valid.

\subsection{Quality factor $M$ as a bridge to possible experimental results.}

So far we have presented our study from the perspective of increasing the angular noise strength at a constant density of obstacles. However, in an experimental setup such a strategy might be not possible. While in experiments it is difficult to control the rotational diffusion coefficient, the obstacle density may be varied. 
For a direct comparison with experimentally accessible control parameters 
we represent the data in terms of two dimensionless quantities, the quality factor $M$ and the  ratio of diffusion in a crowded system  to the diffusion in an obstacle-free system.
We show how the diffusivity is amplified or suppressed at different obstacle densities for a range of quality factors.
We discuss $D=D(R,M,n^*)$ as a function of the quality factor $M$ with fixed orbit radius and obstacle density, rather than as a function of the (reduced) orientational diffusivity. 
Therefore we define the rescaled diffusivity 
\begin{equation}
     \tilde{D}  = \tilde{D}(R,M,n^*):= D(R,M,n^*) / D(R, M, n^* = 0)\,,  
     \label{eq:an_equality_with_M}
\end{equation}
which by construction reduces at zero obstacle density to $\tilde{D}(R,M,n^* = 0) = 1$. 
Indeed, from the simulations we see that for low obstacle density $\tilde{D} \approx 1$ holds for quality factors below the optimal one, $M \lesssim M^\textrm{\scriptsize opt} = 1/2\pi$ [Fig.~\ref{fig:dependence_on_M}(a,b)]. 
At these low densities, increasing the quality factor beyond the optimal one leads to an increase of the rescaled diffusivity.
Moreover,  a larger amplification is observed for a larger $R/\sigma$ ratio.
This observation suggests that scattering at an isolated obstacle is more efficient in accelerating transport if the orbit radius is larger than the obstacle size. Propagation around the boundary of a small obstacle cluster, or even a single obstacle, facilitates meandering thus increasing the diffusivity. This effect becomes more pronounced for larger radii.

For somewhat larger densities, the diffusivity shows a minimum at the optimal quality factor value [Fig.~\ref{fig:dependence_on_M}(c-f)].
This occurs together with another striking observation: for densities $n^* \lesssim 0.25-0.30$ there is an approximate data collapse for different radii $R$ [Fig.~\ref{fig:dependence_on_M}(e,f)].
%
This feature is  a peculiarity for microswimmers and is connected to the wall-following mechanism. 
We have checked that for specular scattering off the obstacles as in magneto-transport this data collapse does not occur. 

In the case of the highest density $n^*=0.35$ close to the percolation transition $n^*_c$  there is still a pronounced minimum in the  amplification at $M=M^{\textrm{\scriptsize opt}}$ [Fig.~\ref{fig:dependence_on_M}(g)], yet the dependence on the orbit radius becomes significant.
In the ideal model at obstacle densities around $n^*=0.35$ the disordered structure ceases to amplify transport and starts to suppress it, as at $n^*_c$ one expects the diffusivity to become zero. At this density the pockets of the void space appear connected by narrow channels. From the data one can conclude that particles with smaller orbit radii are more efficient in moving through such a structure than their counterparts with larger orbit radii. Then the order of the curves at high density [Fig.~\ref{fig:dependence_on_M}(g)] is reversed in comparison to the low density $n^*=0.05$, [Fig.~\ref{fig:dependence_on_M}(a)]. The same effect has been reported for ideal microswimmers where it was also seen that the diffusivity at very high obstacle densities is larger for smaller orbit radii~\cite{Chepizhko2019}, compare aquamarine points in Fig.~\ref{fig:smearing}(b) $R=0.5\sigma$ and (d) $R=2.0\sigma$. 

To summarize this part, we find that at all densities the shape of the amplification curve remains the same. At low $M$ there is no to little amplification. At $M^{\textrm{\scriptsize opt}}$ the amplification is minimal, but then quickly increases for larger values of $M$.

The properties described above can serve as a guideline for future experiments. The quality factor of the orbit can be easily identified for the given experimental probe particle (biological or artificial) and then one can verify the prediction for the amplification curves at different obstacle densities. The best correspondence between experiment and theory is expected for  particles that are well described by our model. This is those that become trapped around obstacles and can travel large distances before departing from the obstacle boundary.

\section{Summary and Conclusions}
We have investigated the transport properties of a noisy circular microswimmer exploring  a heterogeneous environment that consists of overlapping non-permeable obstacles.~We have employed a boundary-following mechanism accounting for the  microswimmer's specific interactions with obstacles. These  interactions are distinct  from non-motile particles which instead exhibit specular reflection. 
For our microswimmers, a small noise and a low spatial disorder lead to an enhancement of transport and to an overall increase of the diffusivity. 
Adding angular noise  to an ideal circle microswimmer allows the active particle to meander from one cluster of overlapping obstacles to another faster and more efficiently. 
Additionally, by increasing the obstacle density an amplification of transport is achieved, since obstacles promote propagation in a swift way along their edges. 
However, a further increase of randomness, by the addition of more angular noise strongly suppresses  transport again, in particular for large orbit radii.

As our main finding we have identified that the time to cover a characteristic inter-obstacle distance is the main parameter that governs the amplification-suppression patterns. We have shown that at small radii the microswimmers with noise can propagate efficiently in a diffusive regime. However, for larger radii the diffusive regime causes the transport suppression.

To identify which effects are purely the consequences of the introduced wall-following mechanism we have performed numerical simulations of a corresponding model but with specular reflections from the obstacles. 
Such a model is relevant for  magneto-transport of electrons in disordered environments  where  noise due to scattering from phonons becomes relevant. 
For such specular scattering  the diffusivity is non-monotonic in the obstacle density  with a pronounced maximum~\cite{Schirmacher2015,Siboni2018} at an intermediate value far from $n^*_c$. 
If the noise is increased, the maximum of the  diffusivity is systematically shifted to even lower obstacle densities, thus at high densities the diffusivity becomes suppressed in strong contrast to the wall-following mechanism.

To relate our model to experiments, more details on the interaction mechanism with the obstacles should be taken into account. 
For example, as a first step one can let the direction of motion evolve by a noisy dynamics while the swimmer interacts with the obstacle. Moreover, the explicit equation for the swimmer interaction with an obstacle could be taken from Ref.~\cite{Spagnolie2015} to substitute the rule in the current paper. These will serve as important steps in further modeling of the microswimmer dynamics in crowded media. We expect the main findings to remain the same, but a better correspondence with experiments could be achieved.

Insight can be taken from recent biological experiments. For example,  \emph{E. Coli} bacteria do not just follow the boundaries, but can experience  specular scattering events depending on the angle of approach to the obstacle~\cite{Makarchuk2019}.
Their diffusivity increases if a small number of obstacles is added, but decreases upon further increase of the obstacle density. This is in qualitative agreement with our results.

If the translational diffusion is non-zero the meandering transition will be smeared even for vanishing rotational diffusion. Correspondingly, one anticipates parameter regimes (for particles closer to passive ones) such that translational diffusion yields the main contribution to the smearing, rather than orientational diffusion. An interesting extension of our work would be to elaborate the competition between both stochastic noises.

It is also interesting to generalize to visco-elastic media~\cite{Lozano2018,Narinder2018}, as many  microorganisms  move in non-Newtonian biological fluids~\cite{Martinez2014,Zoettl2019}. 
This can be  achieved in principle by changing the dynamic rules of  motion in the void space, as well as the particle-obstacle interaction. 
Another extension of the model is to consider  driven systems~\cite{Reichhardt2018JPCM,Leitmann2013,Reichhardt2014,Reichhardt2019} where external driving forces, flows, or chemical gradients are present. 
This extension is important, as  experiments on motile bacteria transport through porous media are typically performed in microfluidic devices with an imposed flow~\cite{Creppy2019}.

A natural extension of our model is to consider interacting active circle swimmers in the presence of obstacles. For the passive counterpart recent simulations~\cite{Schnyder2018} have revealed a striking speed-up of transport, while in bulk interacting particles typically slow down transport and may lead to structural arrest. Active circle swimmers at low densities may similarly promote transport since they will push each other to the walls where the wall-following mechanism sets in. In contrast at high swimmer densities they may get trapped or jammed at the boundaries of the obstacles, such that the wall-following mechanism is no longer efficient.   


For future applications one may ask if an agent can adjust its motility parameters depending on its local environment. In state-of-the-art experiments~\cite{Lavergne2019,FernandezRodriguez2019} a direct control of the active particles can be achieved to mimic such a behavior. In the future active agents may be designed that display a  dynamic feedback to optimize locally their transport in a given landscape of obstacles. A step even further are smart agents that can design their own rules to achieve common goals which have been discussed only recently~\cite{Ried2019,Charlesworth2019}. 

%
%

\section*{Acknowledgements}
We thank Felix H\"ofling for insightful discussions. 
We thank Charlotte Petersen for discussions and careful reading of the final version of the manuscript.
OC is supported by the Austrian Science Fund (FWF): M 2450-NBL. TF acknowledges funding by 
FWF: P 28687-N27. The computational results presented have been achieved in part using the HPC infrastructure LEO of the University of Innsbruck.

%
%
%
%
%

\section*{References}
\bibliographystyle{iopart-num}
\bibliography{microswimmers}

\providecommand{\newblock}{}
\begin{thebibliography}{10}
\expandafter\ifx\csname url\endcsname\relax
  \def\url#1{{\tt #1}}\fi
\expandafter\ifx\csname urlprefix\endcsname\relax\def\urlprefix{URL }\fi
\providecommand{\eprint}[2][]{\url{#2}}

\bibitem{Romanczuk2012}
Romanczuk P, B{\"a}r M, Ebeling W, Lindner B and Schimansky-Geier L 2012 {\em
  The European Physical Journal Special Topics\/} {\bf 202} 1--162 ISSN
  1951-6401 \urlprefix\url{http://dx.doi.org/10.1140/epjst/e2012-01529-y}

\bibitem{Elgeti2015}
Elgeti J, Winkler R~G and Gompper G 2015 {\em Reports on Progress in Physics\/}
  {\bf 78} 056601
  \urlprefix\url{http://stacks.iop.org/0034-4885/78/i=5/a=056601}

\bibitem{BechingerReview2016}
Bechinger C, Di~Leonardo R, L\"owen H, Reichhardt C, Volpe G and Volpe G 2016
  {\em Rev. Mod. Phys.\/} {\bf 88}(4) 045006
  \urlprefix\url{https://link.aps.org/doi/10.1103/RevModPhys.88.045006}

\bibitem{Reichhardt2017review}
Reichhardt C~O and Reichhardt C 2017 {\em Annual Review of Condensed Matter
  Physics\/} {\bf 8} 51--75
  \urlprefix\url{https://doi.org/10.1146/annurev-conmatphys-031016-025522}

\bibitem{Makarchuk2019}
Makarchuk S, Braz V~C, Ara{\'u}jo N~A~M, Ciric L and Volpe G 2019 {\em Nature
  Communications\/} {\bf 10} 4110 ISSN 2041-1723
  \urlprefix\url{https://doi.org/10.1038/s41467-019-12010-1}

\bibitem{Chepizhko2019}
Chepizhko O and Franosch T 2019 {\em Soft Matter\/} {\bf 15}(3) 452--461
  \urlprefix\url{http://dx.doi.org/10.1039/C8SM02030B}

\bibitem{Jakuszeit2019}
Jakuszeit T, Croze O~A and Bell S 2019 {\em Phys. Rev. E\/} {\bf 99}(1) 012610
  \urlprefix\url{https://link.aps.org/doi/10.1103/PhysRevE.99.012610}

\bibitem{Chepizhko2013}
Chepizhko O and Peruani F 2013 {\em Phys. Rev. Lett.\/} {\bf 111}(16) 160604
  \urlprefix\url{http://link.aps.org/doi/10.1103/PhysRevLett.111.160604}

\bibitem{Zeitz2017}
Zeitz M, Wolff K and Stark H 2017 {\em The European Physical Journal E\/} {\bf
  40} 23 ISSN 1292-895X
  \urlprefix\url{http://dx.doi.org/10.1140/epje/i2017-11510-0}

\bibitem{Morin2017pre}
Morin A, Lopes~Cardozo D, Chikkadi V and Bartolo D 2017 {\em Phys. Rev. E\/}
  {\bf 96}(4) 042611
  \urlprefix\url{https://link.aps.org/doi/10.1103/PhysRevE.96.042611}

\bibitem{SosaHernandez2017}
Sosa-Hern\'andez J~E, Santill\'an M and Santana-Solano J 2017 {\em Phys. Rev.
  E\/} {\bf 95}(3) 032404
  \urlprefix\url{https://link.aps.org/doi/10.1103/PhysRevE.95.032404}

\bibitem{Frangipane2019}
Frangipane G, Vizsnyiczai G, Maggi C, Savo R, Sciortino A, Gigan S and
  Di~Leonardo R 2019 {\em Nature Communications\/} {\bf 10} 2442 ISSN 2041-1723
  \urlprefix\url{https://doi.org/10.1038/s41467-019-10455-y}

\bibitem{Benichou2014}
B{\'e}nichou O, Illien P, Oshanin G, Sarracino A and Voituriez R 2014 {\em
  Phys. Rev. Lett.\/} {\bf 113}(26) 268002
  \urlprefix\url{https://link.aps.org/doi/10.1103/PhysRevLett.113.268002}

\bibitem{Reichhardt2018JPCM}
Reichhardt C and Reichhardt C~J~O 2018 {\em Journal of Physics: Condensed
  Matter\/} {\bf 30} 015404
  \urlprefix\url{http://stacks.iop.org/0953-8984/30/i=1/a=015404}

\bibitem{Benichou2018}
B{\'e}nichou O, Illien P, Oshanin G, Sarracino A and Voituriez R 2018 {\em
  Journal of Physics: Condensed Matter\/} {\bf 30} 443001
  \urlprefix\url{http://stacks.iop.org/0953-8984/30/i=44/a=443001}

\bibitem{Peter2018}
P{\'e}ter H, Lib{\'a}l A, Reichhardt C and Reichhardt C~J~O 2018 {\em
  Scientific Reports\/} {\bf 8} 10252 ISSN 2045-2322
  \urlprefix\url{https://doi.org/10.1038/s41598-018-28256-6}

\bibitem{Lozano2018}
Lozano C, Gomez-Solano J~R and Bechinger C 2018 {\em New Journal of Physics\/}
  {\bf 20} 015008 \urlprefix\url{https://doi.org/10.1088/1367-2630/aa9ed1}

\bibitem{Narinder2018}
Narinder N, Bechinger C and Gomez-Solano J~R 2018 {\em Phys. Rev. Lett.\/} {\bf
  121}(7) 078003
  \urlprefix\url{https://link.aps.org/doi/10.1103/PhysRevLett.121.078003}

\bibitem{tenHagen2014}
ten Hagen B, K{\"u}mmel F, Wittkowski R, Takagi D, L{\"o}wen H and Bechinger C
  2014 {\em Nature Communications\/} {\bf 5} 4829 article
  \urlprefix\url{http://dx.doi.org/10.1038/ncomms5829}

\bibitem{Friedrich2008}
Friedrich B~M and J\"ulicher F 2008 {\em New Journal of Physics\/} {\bf 10}
  123025 \urlprefix\url{http://stacks.iop.org/1367-2630/10/i=12/a=123025}

\bibitem{Kaupp2016}
Kaupp U~B and Alvarez L 2016 {\em The European Physical Journal Special
  Topics\/} {\bf 225} 2119--2139 ISSN 1951-6401
  \urlprefix\url{https://doi.org/10.1140/epjst/e2016-60097-1}

\bibitem{BrunCosmeBruny2019}
Brun-Cosme-Bruny M, Bertin E, Coasne B, Peyla P and Rafa\"{i} S 2019 {\em The
  Journal of Chemical Physics\/} {\bf 150} 104901 (\textit{Preprint}
  \eprint{https://doi.org/10.1063/1.5081507})
  \urlprefix\url{https://doi.org/10.1063/1.5081507}

\bibitem{Reichhardt2013}
Reichhardt C and Reichhardt C~J~O 2013 {\em Phys. Rev. E\/} {\bf 88}(4) 042306
  \urlprefix\url{https://link.aps.org/doi/10.1103/PhysRevE.88.042306}

\bibitem{Denissenko2012}
Denissenko P, Kantsler V, Smith D~J and Kirkman-Brown J 2012 {\em Proceedings
  of the National Academy of Sciences\/} {\bf 109} 8007--8010
  \urlprefix\url{https://doi.org/10.1073/pnas.1202934109}

\bibitem{Takagi2014}
Takagi D, Palacci J, Braunschweig A~B, Shelley M~J and Zhang J 2014 {\em Soft
  Matter\/} {\bf 10}(11) 1784--1789
  \urlprefix\url{http://dx.doi.org/10.1039/C3SM52815D}

\bibitem{Nosrati2015}
Nosrati R, Driouchi A, Yip C~M and Sinton D 2015 {\em Nature Communications\/}
  {\bf 6} 8703

\bibitem{Brown2016}
Brown A~T, Vladescu I~D, Dawson A, Vissers T, Schwarz-Linek J, Lintuvuori J~S
  and Poon W~C~K 2016 {\em Soft Matter\/} {\bf 12}(1) 131--140
  \urlprefix\url{http://dx.doi.org/10.1039/C5SM01831E}

\bibitem{Wykes2017}
Davies~Wykes M~S, Zhong X, Tong J, Adachi T, Liu Y, Ristroph L, Ward M~D,
  Shelley M~J and Zhang J 2017 {\em Soft Matter\/} {\bf 13}(27) 4681--4688
  \urlprefix\url{http://dx.doi.org/10.1039/C7SM00203C}

\bibitem{Lauga2006}
Lauga E, DiLuzio W~R, Whitesides G~M and Stone H~A 2006 {\em Biophysical
  Journal\/} {\bf 90} 400--412 ISSN 0006-3495
  \urlprefix\url{http://dx.doi.org/10.1529/biophysj.105.06940}

\bibitem{Berke2008}
Berke A~P, Turner L, Berg H~C and Lauga E 2008 {\em Phys. Rev. Lett.\/} {\bf
  101}(3) 038102
  \urlprefix\url{http://link.aps.org/doi/10.1103/PhysRevLett.101.038102}

\bibitem{Spagnolie2015}
Spagnolie S~E, Moreno-Flores G~R, Bartolo D and Lauga E 2015 {\em Soft
  Matter\/} {\bf 11}(17) 3396--3411
  \urlprefix\url{http://dx.doi.org/10.1039/C4SM02785J}

\bibitem{Kuron2019}
Kuron M, St{\"{a}}rk P, Holm C and de~Graaf J 2019 {\em Soft Matter\/}  Advance
  Article \urlprefix\url{http://dx.doi.org/10.1039/C9SM00692C}

\bibitem{Bertrand2018}
Bertrand T, Zhao Y, B\'enichou O, Tailleur J and Voituriez R 2018 {\em Phys.
  Rev. Lett.\/} {\bf 120}(19) 198103
  \urlprefix\url{https://link.aps.org/doi/10.1103/PhysRevLett.120.198103}

\bibitem{Kamal2018}
Kamal A and Keaveny E~E 2018 {\em Journal of The Royal Society Interface\/}
  {\bf 15} 20180592 \urlprefix\url{https://doi.org/10.1098/rsif.2018.0592}

\bibitem{Creppy2019}
Creppy A, Cl\'ement E, Douarche C, D'Angelo M~V and Auradou H 2019 {\em Phys.
  Rev. Fluids\/} {\bf 4}(1) 013102
  \urlprefix\url{https://link.aps.org/doi/10.1103/PhysRevFluids.4.013102}

\bibitem{Chamolly2017}
Chamolly A, Ishikawa T and Lauga E 2017 {\em New Journal of Physics\/} {\bf 19}
  115001 \urlprefix\url{http://stacks.iop.org/1367-2630/19/i=11/a=115001}

\bibitem{Sandor2017}
S\'andor C, Lib\'al A, Reichhardt C and Reichhardt C~J~O 2017 {\em Phys. Rev.
  E\/} {\bf 95}(1) 012607
  \urlprefix\url{https://link.aps.org/doi/10.1103/PhysRevE.95.012607}

\bibitem{Lorentz1905}
Lorentz H~A 1905 {\em Arch. Neerl. Sci. Exactes Nat.\/} {\bf 10} --

\bibitem{Bauer2010}
Bauer T, H{\"o}f{\/}ling F, Munk T, Frey E and Franosch T 2010 {\em The
  European Physical Journal Special Topics\/} {\bf 189} 103--118 ISSN 1951-6401
  \urlprefix\url{https://doi.org/10.1140/epjst/e2010-01313-1}

\bibitem{Mandal2017}
Mandal S, Spanner-Denzer M, Leitmann S and Franosch T 2017 {\em The European
  Physical Journal Special Topics\/} {\bf 226} 3129--3156 ISSN 1951-6401
  \urlprefix\url{https://doi.org/10.1140/epjst/e2017-70077-5}

\bibitem{Schnyder2015}
Schnyder S~K, Spanner M, H{\"o}f{\/}ling F, Franosch T and Horbach J 2015 {\em
  Soft Matter\/} {\bf 11}(4) 701--711
  \urlprefix\url{http://dx.doi.org/10.1039/C4SM02334J}

\bibitem{Spanner2013}
Spanner M, Schnyder S~K, H\"ofling F, Voigtmann {\relax Th} and Franosch T 2013
  {\em Soft Matter\/} {\bf 9}(5) 1604--1611
  \urlprefix\url{http://dx.doi.org/10.1039/C2SM27060A}

\bibitem{Hofling2013}
H{\"o}f{\/}ling F and Franosch T 2013 {\em Reports on Progress in Physics\/}
  {\bf 76} 046602
  \urlprefix\url{http://stacks.iop.org/0034-4885/76/i=4/a=046602}

\bibitem{Petersen2019}
Petersen C~F and Franosch T 2019 {\em Soft Matter\/} {\bf 15}(19) 3906--3913
  \urlprefix\url{http://dx.doi.org/10.1039/C9SM00442D}

\bibitem{Kummel2013}
K\"ummel F, ten Hagen B, Wittkowski R, Buttinoni I, Eichhorn R, Volpe G,
  L\"owen H and Bechinger C 2013 {\em Phys. Rev. Lett.\/} {\bf 110}(19) 198302
  \urlprefix\url{http://link.aps.org/doi/10.1103/PhysRevLett.110.198302}

\bibitem{Utada2014}
Utada A~S, Bennett R~R, Fong J~C~N, Gibiansky M~L, Yildiz F~H, Golestanian R
  and Wong G~C~L 2014 {\em Nature Communications\/} {\bf 5} 4913 article
  \urlprefix\url{http://dx.doi.org/10.1038/ncomms5913}

\bibitem{PerezIpia2019}
Ipi{\~{n}}a E~P, Otte S, Pontier-Bres R, Czerucka D and Peruani F 2019 {\em
  Nature Physics\/} {\bf 15} 610--615
  \urlprefix\url{https://doi.org/10.1038/s41567-019-0460-5}

\bibitem{vanTeeffelen2008}
van Teeffelen S and L\"owen H 2008 {\em Phys. Rev. E\/} {\bf 78}(2) 020101
  \urlprefix\url{https://link.aps.org/doi/10.1103/PhysRevE.78.020101}

\bibitem{Kurzthaler2017}
Kurzthaler C and Franosch T 2017 {\em Soft Matter\/} {\bf 13}(37) 6396--6406
  \urlprefix\url{http://dx.doi.org/10.1039/C7SM00873B}

\bibitem{Basu2019}
Basu U, Majumdar S~N, Rosso A and Schehr G 2019 {\em Physical Review E\/} {\bf
  100} \urlprefix\url{https://doi.org/10.1103/physreve.100.062116}

\bibitem{Scala2007}
Scala A, Voigtmann {\relax Th} and De~Michele C 2007 {\em The Journal of
  Chemical Physics\/} {\bf 126} 134109 (\textit{Preprint}
  \eprint{https://doi.org/10.1063/1.2719190})
  \urlprefix\url{https://doi.org/10.1063/1.2719190}

\bibitem{Hoefling2006}
H\"ofling F, Franosch T and Frey E 2006 {\em Phys. Rev. Lett.\/} {\bf 96}(16)
  165901 \urlprefix\url{https://link.aps.org/doi/10.1103/PhysRevLett.96.165901}

\bibitem{Ebbens2010}
Ebbens S, Jones R~A~L, Ryan A~J, Golestanian R and Howse J~R 2010 {\em Phys.
  Rev. E\/} {\bf 82}(1) 015304
  \urlprefix\url{https://link.aps.org/doi/10.1103/PhysRevE.82.015304}

\bibitem{Kuzmany1998}
Kuzmany A and Spohn H 1998 {\em Phys. Rev. E\/} {\bf 57}(5) 5544--5553
  \urlprefix\url{https://link.aps.org/doi/10.1103/PhysRevE.57.5544}

\bibitem{Schirmacher2015}
Schirmacher W, Fuchs B, H\"ofling F and Franosch T 2015 {\em Phys. Rev.
  Lett.\/} {\bf 115}(24) 240602
  \urlprefix\url{http://link.aps.org/doi/10.1103/PhysRevLett.115.240602}

\bibitem{Franosch2010}
Franosch T, H\"ofling F, Bauer T and Frey E 2010 {\em Chemical Physics\/} {\bf
  375} 540 -- 547
  \urlprefix\url{http://www.sciencedirect.com/science/article/pii/S0301010410001898}

\bibitem{Siboni2018}
Siboni N~H, Schluck J, Pierz K, Schumacher H~W, Kazazis D, Horbach J and
  Heinzel T 2018 {\em Phys. Rev. Lett.\/} {\bf 120}(5) 056601
  \urlprefix\url{https://link.aps.org/doi/10.1103/PhysRevLett.120.056601}

\bibitem{Martinez2014}
Martinez V~A, Schwarz-Linek J, Reufer M, Wilson L~G, Morozov A~N and Poon W~C~K
  2014 {\em Proceedings of the National Academy of Sciences\/} {\bf 111}
  17771--17776 ISSN 0027-8424 (\textit{Preprint}
  \eprint{https://www.pnas.org/content/111/50/17771.full.pdf})
  \urlprefix\url{https://www.pnas.org/content/111/50/17771}

\bibitem{Zoettl2019}
Z\"{o}ttl A and Yeomans J~M 2019 {\em Nature Physics\/} {\bf 15} 554--558
  \urlprefix\url{https://doi.org/10.1038/s41567-019-0454-3}

\bibitem{Leitmann2013}
Leitmann S and Franosch T 2013 {\em Phys. Rev. Lett.\/} {\bf 111}(19) 190603
  \urlprefix\url{https://link.aps.org/doi/10.1103/PhysRevLett.111.190603}

\bibitem{Reichhardt2014}
Reichhardt C and Olson~Reichhardt C~J 2014 {\em Phys. Rev. E\/} {\bf 90}(1)
  012701 \urlprefix\url{https://link.aps.org/doi/10.1103/PhysRevE.90.012701}

\bibitem{Reichhardt2019}
Reichhardt C and Reichhardt C~J~O 2019 {\em Phys. Rev. E\/} {\bf 100}(1) 012604
  \urlprefix\url{https://link.aps.org/doi/10.1103/PhysRevE.100.012604}

\bibitem{Schnyder2018}
Schnyder S~K and Horbach J 2018 {\em Physical Review Letters\/} {\bf 120}
  \urlprefix\url{https://doi.org/10.1103/physrevlett.120.078001}

\bibitem{Lavergne2019}
Lavergne F~A, Wendehenne H, B\"{a}uerle T and Bechinger C 2019 {\em Science\/}
  {\bf 364} 70--74 \urlprefix\url{https://doi.org/10.1126/science.aau5347}

\bibitem{FernandezRodriguez2019}
Fernandez-Rodriguez M~A, Grillo F, Alvarez L, Rathlef M, Buttinoni I, Volpe G
  and Isa L 2019 Active colloids with position-dependent rotational diffusivity
  (\textit{Preprint} \eprint{arXiv:1911.02291})

\bibitem{Ried2019}
Ried K, M\"{u}ller T and Briegel H~J 2019 {\em {PLOS} {ONE}\/} {\bf 14}
  e0212044 \urlprefix\url{https://doi.org/10.1371/journal.pone.0212044}

\bibitem{Charlesworth2019}
Charlesworth H~J and Turner M~S 2019 {\em Proceedings of the National Academy
  of Sciences\/} {\bf 116} 15362--15367
  \urlprefix\url{https://doi.org/10.1073/pnas.1822069116}

\end{thebibliography}

\end{document}